\documentclass[pra,a4paper,showpacs,superscriptaddress]{revtex4-2}
\usepackage{amsmath}
\usepackage{amsfonts}
\usepackage{graphicx}
\usepackage{longtable}
\newcommand{\be}{\begin{equation}}
\newcommand{\ee}{\end{equation}}

\newcommand{\ba}[1]{\left(\begin{array}{#1}}
\newcommand{\ea}{\end{array}\right)}
\begin{document}

\title{Entanglement and volume monogamy features of permutation symmetric $N$-qubit pure states with $N$-distinct spinors:  GHZ and  W$\bar{\rm W}$ states}

\author{Sudha} 
\affiliation{Department of Physics, Kuvempu University, 
	Shankaraghatta-577 451, Karnataka, India}
\email{tthdrs@gmail.com}
\affiliation{Inspire Institute Inc., Alexandria, Virginia, 22303, USA.}

\author{A. R. Usha Devi} 
\affiliation{Department of Physics, Bangalore University, 
	Bangalore-560 056, India}
\email{ushadevi@bub.ernet.in}
\affiliation{Inspire Institute Inc., Alexandria, Virginia, 22303, USA.}
\email{ushadevi@bub.ernet.in}

\author{H. Akshata Shenoy} 
\affiliation{International Centre for Theory of Quantum Technologies, University of Gd{\'a}nsk, Gd{\'a}nsk, Poland}
\email{akshata.shenoy@ug.edu.pl}

\author{H. S. Karthik} 
\affiliation{International Centre for Theory of Quantum Technologies, University of Gd{\'a}nsk, Gd{\'a}nsk, Poland}
\email{karthik.hs@ug.edu.pl} 

\author{Talath Humera}
\affiliation{Department of Physics, Bangalore University, 
	Bangalore-560 056, India}
\email{talathumera45@gmail.com}

\author{B. P. Govindaraja}
\affiliation{Department of Physics, Kuvempu University, 
	Shankaraghatta-577 451, Karnataka, India}
\email{govindarajabp@gmail.com}
	
\author{A. K. Rajagopal} 
\affiliation{Inspire Institute Inc., McLean, VA 22101, USA.}
\email{attipat.rajagopal@gmail.com}

\date{\today}

\begin{abstract} 
We explore the entanglement features of pure symmetric $N$-qubit states characterized by  $N$-distinct spinors with a particular focus on the Greenberger-Horne-Zeilinger(GHZ) states and W$\bar{\rm W}$, an equal superposition of W and obverse W states. 
Along with a comparison of pairwise entanglement and monogamy properties, we explore the geometric information contained in them by constructing their canonical steering ellipsoids. We obtain the volume monogamy relations satisfied by $\rm{W\bar W}$ states as a function of number of qubits and compare with the maximal monogamy property of GHZ states.
\end{abstract}
\pacs{03.67.Mn, 03.67.-a} 
\maketitle

\section{Introduction} 
Permutation symmetric multiqubit states form an important class among quantum states due to their experimental significance and mathematical elegance~\cite{sym1,sym1a,sym1b,sym2,sym3}. The well-known Greenberger-Horne-Zeilinger(GHZ)~\cite{ghz}, W, and Dicke states~\cite{dicke} etc., belong to this class.
Mathematical simplicity in addressing pure symmetric $N$-qubit states  is owing to the fact that they are confined to the $N+1$ dimensional subspace of the $2^N$ dimensional Hilbert space. The $N+1$ dimensional subspace is the maximal multiplicity space of the collective angular momentum space of $N$-qubits with Dicke states~\cite{dicke}, the common eigenstates of the squared collective angular momentum operator $J^2$ and its $z$-component $J_z$ forming its basis.  
In 1932  Majorana~\cite{majorana} proposed an elegant geometrical visualization for pure symmetric $N$-qubit states as a constellation of $N$-points on the Bloch sphere $S^2$. The representation of pure symmetric multiqubit states in terms of constitutent $N$-qubits (spinors) is called {\em Majorana representation}~\cite{majorana}. Majorana geometric representation has found several significant applications in quantum information processing~\cite{solano,bastin,arus,markham1}.  

Quantum steering ellipsoid~\cite{jevtic2014} offers a novel geometric picturization of a two-qubit states and is useful in understanding quantum correlations such as non-locality, entanglement~\cite{shi2011,MilneNJP2014,MilnePRA2014} and quantum discord~\cite{shi2011,shi}.  The set of all Bloch vectors to which one of the qubits of a two-qubit system can be `steered' when all possible measurements are carried out on the other qubit correspond to {\emph{quantum steering ellipsoid}}~\cite{jevtic2014}. It has been identified that the volume of the steering ellipsoids~\cite{jevtic2014} corresponding to the two-qubit subsystems of an $N$-qubit state, $N\geq 3$, effectively captures monogamy properties of the state~\cite{MilneNJP2014,MilnePRA2016}. Milne et.al ~\cite{MilneNJP2014} proposed a monogamy relation, in terms of the volumes of the quantum steering ellipsoids of two-qubit subsystems of a $3$-{\emph{qubit pure state}} which is stricter than the Coffman-Kundu-Wootters(CKW) monogamy relation~\cite{ckw}. A volume monogamy relation satisfied by pure as well as mixed $N$-qubit states has been obtained in ~\cite{MilnePRA2016} and is helpful in quantifying the shareability properties of the $N$-qubit state. 

The steering ellipsoid of a two-qubit state that has attained a canonical form under suitable local operations on {\emph {both the qubits}} is the so-called {\emph{canonical steering ellipsoid}}~\cite{verstraete2001,fvthesis,supra} and provides another geometric representation of a two-qubit state. The canonical steering ellipsoid of any two-qubit state is shown to have only two distinct forms~\cite{supra} and  provide a much simpler geometric picture representing two-qubit states. 

The volume monogamy relations for permutation symmetric $3$-qubit pure states with two and three distinct spinors are established in \cite{can} using the features of their respective steering ellipsoids. The canonical steering ellipsoids of the entire class of permutation symmetric $N$-qubit states with two distinct spinors are obtained in ~\cite{pra23} and the nature of the volume monogamy relation with increasing $N$ is analyzed~\cite{pra23}. 
In addition, the obesity of the steering ellipsoids is made use of to obtain expressions for concurrence of the two-qubit subsystems of the $N$-qubit states under consideration~\cite{can,pra23}. In this paper, we construct the canonical steering ellipsoids of the $N$-qubit GHZ and W$\bar{\rm W}$ states and analyze the volume monogamy relations satisfied by them. 

Contents of this paper are organized as follows: In Sec.~II, following a brief overview on   
Majorana  representation~\cite{majorana,solano,bastin,arus} of pure permutation symmetric multiqubit states, we obtain the nature of $N$ distinct spinors characterizing GHZ, W$\bar{\rm W}$ states and their generalized counterparts. Using these, we show that $3$-qubit GHZ and W$\bar{\rm W}$ states are interconvertible under local operations on each qubit. In Sec.~III we analyze the pairwise entanglement features as well as monogamous nature of $N$-qubit GHZ and W$\bar{\rm W}$ states. Following a primer on canonical forms of two-qubit subsystems of pure $N$-qubit state in Sec.~IV, we construct the canonical steering ellipsoids of GHZ and W$\bar{\rm W}$ states in Sec. V. The nature of the volume monogamy relation satisfied by W$\bar{\rm W}$ states is obtained and a comparison with that of GHZ and W-class of states is carried out in Sec. VI.  Concluding remarks are given in Sec. VII.   
  
\section{ Majorana representation of pure symmetric multiqubit states} 
A system of $N$-qubits obeying exchange symmetry gets restricted to a $(N+1)$ dimensional Hilbert space spanned by the basis vectors $\{\vert N/2,k- N/2\rangle, k=0,1,2,
\ldots N \}$ where,    
\begin{eqnarray}
\label{Dicke} 
\vert N/2, k-N/2\rangle &=&\frac{1}{\sqrt{^N C_k}}\,[\vert \underbrace{0, 0, \ldots}_{k\ {\rm times}}\, ,  
\underbrace{1, 1, \ldots}_{N-k\ {\rm times}}\rangle \nonumber \\ 
&&  \ \ \ \ \ \ \ \ \ \ +\ {\rm Permutations}\ ] 
\end{eqnarray}
are the $N+1$ Dicke  states -- expressed in the standard qubit basis $\vert 0\rangle,\ \vert 1\rangle$. 
 
An arbitrary pure symmetric state,    
\begin{equation}
\label{sympure1} 
\vert \Psi_{\rm sym}\rangle =\sum_{k=0}^{N}\, d_k\, \vert N/2, k-N/2\rangle,
\end{equation}
is specified by the $(N+1)$ complex coefficients $d_k.$ Eliminating an overall phase and normalizing  the 
state (i.e., $\sum_{k=0}^{N}\, \vert d_k\vert^2=1$) implies that $N$ complex parameters are required to 
completely characterize a pure symmetric state of $N$ qubits.  

Alternately, Majorana~\cite{majorana} expressed the  pure state $\vert \Psi_{\rm sym}\rangle$ as a superposition
of  {\em symmetrized} states of  $N$ spin-$1/2$ particles: 
\begin{eqnarray}
\label{Maj}
& & \vert \Psi_{\rm sym}\rangle={\cal N}\, \sum_{P}\, \hat{P}\, \{\vert \epsilon_1\epsilon_2, 
\ldots  \epsilon_N \rangle\}  
\end{eqnarray}
where
\begin{equation}
\label{Ms}
 \vert \epsilon_s\rangle=\cos(\beta_s/2)\, e^{-i\alpha_s/2}\, \vert 0\rangle
+\sin(\beta_s/2)\, e^{i\alpha_s/2}\, \vert 1\rangle, \ s=1,2,\ldots, N
\end{equation} 
denote spinors constituting the pure symmetric state $\vert \Psi_{\rm sym}\rangle$. Here $\hat{P}$ denotes the set of all $N!$ 
permutations and ${\cal N}$ corresponds to an overall normalization factor. The $N$ complex parameters 
$z_s=\tan\frac{\beta_s}{2}e^{i\alpha_s}$, where $(\alpha_s$, $\beta_s)$ correspond to orientations of the spinor $\vert\epsilon_s\rangle$ (see (\ref{Ms})), offer an alternate parametrization for the pure symmetric $N$ qubit  state $\vert \Psi_{\rm sym}\rangle$. The two representations (\ref{sympure1}) and (\ref{Maj}) of $\vert \Psi_{\rm sym}\rangle$ together lead to the so-called {\emph{Majorana polynomial equation}}:~\cite{arus}
\begin{eqnarray}
\label{cmj02}
 P(z)=\sum_{k=0}^N (-1)^k\,\sqrt{^N\, C_k}\,  d_k\,  \, z^{k}&=&0.
\end{eqnarray} 
\begin{itemize}
\item The solutions $$z_s=\tan\frac{\beta_s}{2}e^{i\alpha_s}$$ of the Majorana polynomial equation (\ref{cmj02}) determine the orientations $(\alpha_s,\beta_s)$ of the  spinors constituting the  state $\vert \Psi_{\rm sym}\rangle$, in terms of the collective parameters $d_k$ (see (\ref{sympure1})). 
\item When the Majorana Polynomial $P(z)$ is of degree $r<N$, it is necessary to recast the polynomial $P(z)$ in terms of $z'=\frac{1}{z}=\cot \left(\frac{\beta_s}{2}\right)\,e^{-i\alpha_s}$ so that the $N-r$ solutions    
\begin{equation} 
\label{Mp2}
P(z')=\sum_{k=0}^N\, (-1)^{N-k}\, \sqrt{^N\, C_k}\,  d_{N-k}\,  \, z'^{N-k}=0
\end{equation}
of (\ref{Mp2}) determine the orientations of the remaining $N-r$ spinors constituting the state $\vert \Psi_{\rm sym}\rangle$. In other words, given the parameters $d_k$, the $N$ roots $z_s, s=1,2,\ldots N$ of the Majorana polynomials (\ref{cmj02}), (\ref{Mp2}) determine the orientations $(\alpha_s,\beta_s)$ of the  spinors 
constituting the pure symmetric $N$-qubit state $\vert \Psi_{\rm sym}\rangle$.  
\item When all the spinors in $(\ref{Maj})$ are distinct, the family of states is denoted by ${\cal D}_{1,\,1,\,1,\ldots,1}$ indicating that each of the $N$-spinors in 
the $N$-qubit pure symmetric state appear only once in the symmetrized combination (\ref{Maj}). The family of states ${\cal D}_{N-k,k}$ denotes the family of states with two distinct spinors one of them repeating $k$ times and the other $N-k$ times in (\ref{Maj}), $k=1,\,2,\,3,\ldots \left[\frac{N}{2}\right]$. 
\item Dicke states are prominent members of the family ${\cal D}_{N-k,k}$. In the following, we show that $N$-qubit GHZ and W$\bar{\rm W}$ states belong to the family ${\cal D}_{1,\,1,\,1,\ldots,1}$, with $N$-distinct spinors. 
\end{itemize}

 \subsection{Majorana spinors of  $N$-qubit GHZ and W$\bar{\rm W}$ states:} 

Consider the $N$-qubit GHZ state 
\begin{equation}
\label{ghznq}
\vert \rm{GHZ}\rangle_N=\frac{\vert 0_10_2\dots 0_N\rangle
+ \vert 1_11_2\dots 1_N\rangle}{\sqrt{2}},  
\end{equation}
expressed in terms of the angular momentum states $\vert jm\rangle$, 
$j=N/2$, $m=-j$ to $j$ as 
\begin{equation}
\label{ghzang}
\vert \rm{GHZ}\rangle_N=\frac{\left\vert \frac{N}{2},\frac{N}{2}\right\rangle
+ \left\vert \frac{N}{2},-\frac{N}{2}\right\rangle}{\sqrt{2}}.  
\end{equation} 
Comparing (\ref{ghzang}) with (\ref{Maj}), it can be seen that there are only two non-zero coefficients $d_0=d_{N}=1/\sqrt{2}$. The Majorana polynomial equation (\ref{cmj02}) for $\vert \rm{GHZ}\rangle_N$ turns out to be 
\begin{eqnarray}
\label{pghz}
1+(-1)^{N}\,z^{N}=0 
\end{eqnarray} 

\begin{itemize}
\item Thus the $N^{\rm th}$ roots of unity determine the $N$-distinct spinors of $\vert \rm{GHZ}\rangle_N$ when $N$ is odd and $N^{\rm th}$ roots of $-1$ when $N$ is even.

\item When $N=3$,  $z_1=\omega^3=1$, $z_2=\omega^2$, $z_3=\omega$ where $\omega=\exp {i\,\pi/3}$ are the cube roots of unity and the Majorana spinors 
\begin{equation}
\label{ghzsp}
\vert\epsilon_1\rangle=\frac{1}{\sqrt{2}}\left(\vert 0\rangle+\vert 1\rangle \right), \ \ 
\vert\epsilon_2\rangle=\frac{1}{\sqrt{2}}\left(\vert 0\rangle+\omega^2\vert 1\rangle \right), \ \ 
\vert\epsilon_3\rangle=\frac{1}{\sqrt{2}}\left(\vert 0\rangle+\omega\vert 1\rangle \right). 
 \end{equation}  
constitute $\vert {\rm GHZ}\rangle_3$. 
\item One may verify explicitly that symmetrization of the spinors (\ref{ghzsp}) leads to the GHZ state:  
\begin{eqnarray}
\label{symghz}
\vert {\rm GHZ}\rangle_3&=&\frac{1}{\sqrt{6}}\left[\vert \epsilon_1, \epsilon_2,  \epsilon_3\rangle+
\vert \epsilon_3, \epsilon_1,  \epsilon_2\rangle  
+\vert \epsilon_2, \epsilon_3,  \epsilon_1\rangle\right. \nonumber \\
&& \left. + \vert \epsilon_2, \epsilon_1,  \epsilon_3\rangle +\vert \epsilon_3, \epsilon_2,  \epsilon_1\rangle+  
+\vert \epsilon_1, \epsilon_3,  \epsilon_2\rangle\right] \nonumber \\ 
& & \nonumber \\
&=&\frac{\vert 0_10_20_3\rangle+\vert 1_11_21_3\rangle}{\sqrt{2}} 
\end{eqnarray}

\item In a similar manner, the fourth roots of $-1$ lead to the following  spinors corresponding to   $\vert {\rm GHZ} \rangle_4$:  
\begin{eqnarray}
\label{ghzspinor4}
\vert \epsilon_1\rangle&=&\frac{1}{\sqrt{2}}\left(\vert 0\rangle+e^{i\pi/4}\vert 1\rangle \right)\nonumber \\
\vert\epsilon_2\rangle&=&\frac{1}{\sqrt{2}}\left(\vert 0\rangle+e^{3i\pi/4}\vert 1\rangle \right)\nonumber \\
\vert\epsilon_3\rangle&=&\frac{1}{\sqrt{2}}\left(\vert 0\rangle+e^{5i\pi/4}\vert 1\rangle \right)\nonumber \\
\vert \epsilon_4\rangle&=&\frac{1}{\sqrt{2}}\left(\vert 0\rangle+e^{7i\pi/4}\vert 1\rangle \right).
\end{eqnarray} 
The symmetrization of the four spinors (\ref{ghzspinor4}) as in (\ref{Maj}) results in 
\begin{eqnarray}
\label{ghzspinor5}
\vert {\rm GHZ}\rangle_4&=&{\cal N}\,  \sum_{P}\, \hat{P}\, \{\vert \epsilon_1, \epsilon_2, 
\epsilon_3,  \epsilon_4 \rangle\}  \nonumber \\
&=&   \frac{\vert 0_10_20_30_4 \rangle+\vert 1_11_21_31_4 \rangle}{{\sqrt{2}}},  
\end{eqnarray}
the $4$-qubit GHZ state expressed in the qubit basis. 
\end{itemize}

It is evident from the discussion above, that the $N$-qubit GHZ state is a pure symmetric state characterized by 
$N$-distinct spinors. In the following, we show that the superposition of $N$-qubit W state $\vert \rm W\rangle_N$  and its obverse state $\vert \rm \bar W\rangle_N$ is a pure symmetric state with $N$ distinct spinors $\vert \epsilon'_r \rangle$, $r=1,\,2,\ldots\,,N$. 

We first express the $N$-qubit W state $\vert \rm{W}\rangle_N$, its obverse state $\vert \rm{\bar W}\rangle_N$ in the angular momentum and qubit basis respectively:
\begin{eqnarray*}
\label{wna}
\vert \rm{W}\rangle_N&=& \left\vert \frac{N}{2},\frac{N}{2}-1\right\rangle 
= \frac{\vert 1_10_2\dots 0_N\rangle+\vert 0_11_20_3\dots 0_N\rangle+\ldots
+ \vert 0_10_2\dots 1_N\rangle}{\sqrt{N}},  \\ 
\label{wbna}
\vert \rm{\bar W}\rangle_N&=& \left\vert \frac{N}{2},1-\frac{N}{2}\right\rangle 
= \frac{\vert 0_11_2\dots 1_N\rangle+\vert 1_10_21_3\dots  1_N\rangle+\ldots
+ \vert 1_11_2\dots 0_N\rangle}{\sqrt{N}},  
\end{eqnarray*} 
The equal superposition of $\vert \rm{W}\rangle_N$, $\vert \rm{\bar W}\rangle_N$, which we refer to as W$\bar{\rm W}$ state, is  given by 
\begin{eqnarray}
\label{wwbna}
\vert \rm{W\bar W}\rangle_N&=& \frac{\left\vert \frac{N}{2},\frac{N}{2}-1\right\rangle+\left\vert \frac{N}{2},1-\frac{N}{2}\right\rangle}{\sqrt{2}} \\ 
\label{wwbnq}
&=& \frac{1}{\sqrt{2}}\left(\left[\frac{\vert 1_10_2\dots 0_N\rangle+\vert 0_11_20_3\dots 0_N\rangle+\ldots
+ \vert 0_10_2\dots 1_N\rangle}{\sqrt{N}}\right] \right. \nonumber \\ 
&+& \left.\left[\frac{\vert 0_11_2\dots 1_N\rangle+\vert 1_10_21_3\dots  1_N\rangle+\ldots
+ \vert 1_11_2\dots 0_N\rangle}{\sqrt{N}}\right] \right).
\end{eqnarray} 
\begin{itemize}
\item Comparing (\ref{wwbna}) with (\ref{sympure1}), we have $d_1=d_{N-1}=1/\sqrt{2}$ as the only non-zero coefficients and hence the corresponding Majorana polynomial equation (\ref{cmj02}) turns out to be
\begin{eqnarray}
\label{pww}
z+(-1)^{N}\,z^{N-1}=0 \Longrightarrow z=0, \ \  1+(-1)^{N}\,z^{N-2}=0.
\end{eqnarray} 
\item The solutions of (\ref{pww}) determine the $N-1$ spinors corresponding to $\vert {\rm W}\bar{\rm W}\rangle_N$ and the solution $z'=1/z=0$ of the Majorana polynomial equation (\ref{Mp2}) determines its $N^{th}$ spinor. Recalling that $z=\tan\frac{\beta}{2}\,e^{i\, \alpha}$, $z'=\cot\frac{\beta}{2}\,e^{-i\, \alpha}$ we proceed to determine the nature of Majorana spinors (see (\ref{Ms})) constituting  $\vert {\rm W}\bar{\rm W}\rangle_N$: 
\begin{eqnarray} 
\label{12}
&&z_1= \tan\frac{\beta_1}{2}\,e^{i\, \alpha_1}=0 \Longrightarrow \beta_1=0,\ \  \alpha_1 \ \mbox{arbitrary}. 
\end{eqnarray} 
Let us choose $\alpha_1=0$ to obtain 
\begin{equation}
\label{wwf}
\vert\epsilon'_1\rangle=\vert 0\rangle.
\end{equation} 
\item The other $N-2$ spinors  of the  $\vert {\rm W}\bar{\rm W}\rangle_N$ state are then given by 
\begin{eqnarray} 
\label{zwo}
z^{N-2}=1 \ \ &\mbox{when}& \ \  \ \ N\  \mbox{is odd} \\ 
\label{zwe}
z^{N-2}=-1 \ \ &\mbox{when}& \ \  \ \ N \ \mbox{is even} \ \
\end{eqnarray}
\item Note that the $N^{\rm th}$ spinor corresponds to the solution $z'_N=\cot\frac{\beta_N}{2}\,e^{-i\, \alpha_N}=0$ of (\ref{Mp2}) and we get 
$\beta_N=\pi$, $\alpha_N$ arbitrary. Choosing  $\alpha_N=0$
 we find that 
\begin{equation}
\label{wwfa}
\vert \epsilon'_N\rangle=\vert 1 \rangle.
\end{equation}
\item In adition to the two spinors $\vert \epsilon'_1\rangle=\vert 0\rangle$, $\vert\epsilon'_N\rangle=\vert 1\rangle$ (which are irrespective of any $N$) constituting the state $\vert {\rm W}\bar{\rm W}\rangle_N$, rest of the  distinct spinors are obtained to be   
the $(N-2)^{\rm th}$ roots of unity when $N$ is odd and $(N-2)^{\rm th}$  roots of $-1$ when $N$ is even~\cite{fn}. 

\item For $N=3$, the spinor corresponding to the solution $z=1=\tan\frac{\beta_2}{2}\,e^{i\, \alpha_2}$ of (\ref{zwo}) 
is characterized by the parameters $\beta_2=\pi/2$ and $\alpha_2=0$. In other words,  the {\em three} distinct spinors $\vert \epsilon'_r\rangle$, $r=1,\,2,\,3$ constituting $\vert {\rm W\bar W} \rangle_3$ are given by  
\begin{equation}
\label{wsp}
\vert\epsilon'_1\rangle= \vert 0\rangle, \   \mbox{and} \ \ 
\vert\epsilon'_2\rangle=\frac{\vert 0\rangle+i\vert 1\rangle}{\sqrt{2}}, \ \ 
\vert\epsilon'_3\rangle=\vert 1\rangle.
 \end{equation}
\item When $N=4$, the Majorana spinors corresponding to $\vert {\rm W\bar W} \rangle_4$ are given by  
\begin{eqnarray}
\label{n4wwbsp}
\vert \epsilon'_1\rangle&=&\vert 0\rangle, \nonumber \\
\vert \epsilon'_2\rangle&=&\frac{1}{\sqrt{2}}\left(\vert 0\rangle+i\vert 1\rangle \right)\nonumber \\
\vert \epsilon'_3\rangle&=&\frac{1}{\sqrt{2}}\left(\vert 0\rangle-i\vert 1\rangle \right) \\ 
\vert \epsilon'_4\rangle&=&\vert 1\rangle. \nonumber
\end{eqnarray}
\item One may verify explicitly that 
\begin{eqnarray}
\label{symww}
\vert W\bar W \rangle_3&=&\frac{1}{\sqrt{6}}\left[\vert \epsilon'_1, \epsilon'_2,  \epsilon'_3\rangle+
\vert \epsilon'_3, \epsilon'_1,  \epsilon'_2\rangle  
+\vert \epsilon'_2, \epsilon'_3,  \epsilon'_1\rangle
+ \vert \epsilon'_2, \epsilon'_1,  \epsilon'_3\rangle +\vert \epsilon'_3, \epsilon'_2,  
\epsilon'_1\rangle+  
+\vert \epsilon'_1, \epsilon'_3,  \epsilon'_2\rangle\right] \nonumber \\
& & \nonumber \\
&=&\frac{\vert 0_10_21_3\rangle+\vert 0_1 1_2 0_3\rangle+\vert 1_1 0_2 0_3\rangle+\vert 1_1 1_2 0_3\rangle+\vert 1_10_21_3\rangle+\vert 0_11_21_3\rangle}{\sqrt{6}}.
 \end{eqnarray}
Similarly one may  verify explicitly that  $\vert {\rm W\bar W} \rangle_4$ is constructed by symmetrizing the 4 spinors given in  (\ref{n4wwbsp}). 
\item From the above discussion it is clear that both  $\vert \rm{GHZ}\rangle_N$ and $\vert {\rm W}\bar{\rm W}\rangle_N$ belong to the family ${\cal D}_{1,1,1,\ldots,1}$ of pure symmetric $N$-qubit states characterized by $N$-distinct spinors.   
\end{itemize}

\subsection{Majorana spinors of generalised $N$-qubit GHZ and W$\bar{\rm W}$  states} 
Generalized  $N$-qubit GHZ and W$\bar{\rm W}$ states are respectively given by
 \begin{eqnarray} 
\label{ghzg}  
\vert {\rm GHZ}\rangle_N^{\rm gen} &=& \cos \frac{\theta}{2}\left\vert\frac{N}{2},\frac{N}{2} \right\rangle + 
\sin \frac{\theta}{2}\,\left\vert\frac{N}{2},-\frac{N}{2} \right\rangle
\end{eqnarray}
and  
\begin{eqnarray}
\label{wwbg}
\vert {\rm{W\bar W}}\rangle_N^{\rm gen} &=& \cos \frac{\theta}{2}\left\vert\frac{N}{2},\frac{N}{2}-1 \right\rangle + 
\sin \frac{\theta}{2}\,\left\vert\frac{N}{2},1-\frac{N}{2} \right\rangle.
\end{eqnarray}
where $\theta\in(0,\pi/4).$

\begin{itemize} 
\item From (\ref{sympure1}), (\ref{ghzg}), (\ref{wwbg}), we have $d_0=\sin (\theta/2)$, $d_{N}=\cos (\theta/2)$ for $\vert {\rm GHZ}\rangle_N^{\rm gen}$; 
$d_1=\sin (\theta/2)$ and $d_{N-1}=\cos (\theta/2)$ for $\vert {\rm W\bar W}\rangle_N^{\rm gen}$. 
\item For the $3$-qubit state $\vert {\rm GHZ}\rangle_3^{\rm gen}$, the Majorana polynomial equation (\ref{cmj02}) takes the form 
\[
\sin \frac{\theta}{2}- z^3\,\cos \frac{\theta}{2}=0 \Longrightarrow 1 - \,\left( \frac{z}{\eta}\right)^3=0\ \ \mbox{where} \ \ \eta=\left(\tan \frac{\theta}{2}\right)^{\frac{1}{3}}
\]
 leading to the roots 
\begin{equation}
 z_1\, =\eta\,\omega^3=\eta, \ \ z_2= \eta\omega^2,\ \  z_3=\eta\omega\ \ 
\end{equation} 
$\omega, \omega^2,\omega^3=1$ are the cube roots of unity. 
\item The three Majorana spinors corresponding to $\vert {\rm GHZ}\rangle_3^{\rm gen}$ 
(see (\ref{ghzg})) are therefore given by  
\begin{equation}
\label{ghzgsp}
\vert\epsilon_1\rangle^{\rm gen}=\frac{1}{\sqrt{1+\eta^2}}\, \left(\begin{array}{c}1\\
 \eta \end{array}\right),\ \ 
\vert\epsilon_2\rangle^{\rm gen}=\frac{1}{\sqrt{1+\eta^2}}\, \left(\begin{array}{c}1\\
\eta\omega\, \end{array}\right),\ \
\vert\epsilon_3\rangle^{\rm gen}=\frac{1}{\sqrt{1+\eta^2}}\, \left(\begin{array}{c}1\\
 \eta \omega^2\end{array}\right).
 \end{equation}

\item With $d_{1}=\sin (\theta/2)$ and $d_{2}=\cos (\theta/2)$ for $\vert {\rm W\bar W}\rangle_3^{\rm gen}$, the independent roots of Majorana polynomial equation (see (\ref{cmj02})) $z\,\sin\frac{\theta}{2}-z^2\cos \frac{\theta}{2}=0$
and  (see (\ref{Mp2}))
$z'{^{2}}\,\cos\frac{\theta}{2}-z'\sin \frac{\theta}{2}\,=0$,   $z'=1/z$
are 
\begin{equation}
 z_1=0,\ \ z_2=\tan \frac{\theta}{2}=\eta^3, \ \ z_3=\frac{1}{z}=0.
\end{equation}
\item Thus the Majorana spinors of $\vert {\rm W\bar W}\rangle_3^{\rm gen}$ turn out be 
\be
\label{wwbgsp}
\vert\epsilon'_1\rangle^{\rm gen}= \left(\begin{array}{c}1\\
0\end{array}\right), \ \ 
\vert\epsilon'_2\rangle^{\rm gen}=\frac{1}{\sqrt{1+\eta^6}}\,  \left(\begin{array}{c}1\\
\eta^3 \, \end{array}\right),\ \ \\ 
\vert\epsilon'_3\rangle^{\rm gen}= \left(\begin{array}{c}0\\
1\, \end{array}\right).\ 
 \ee 
\item In general, the $N$ spinors corresponding to $\vert {\rm GHZ} \rangle_N^{\rm gen}$ are  given by 
\begin{equation}
\label{ghzgspN}
\vert\epsilon_k\rangle=\frac{1}{\sqrt{1+\chi^2}}\, \left(\begin{array}{c}1\\
 \chi e^{i(\frac{\phi+2k\pi}{N})}
\end{array}\right),\ \  k=0,\,1,\,2,\ldots\,N-1, \ \ \chi=\left(\tan \frac{\theta}{2}\right)^{\frac{1}{N}}
\end{equation}.
Note that that $\phi=0$ when $N$ is odd and $\phi=\pi$ when $N$ is even. 

\item Similarly the $N$ spinors corresponding to  $\vert {\rm W\bar W} \rangle_N^{\rm gen}$ are, 
\begin{eqnarray}
\label{wspN}
\vert\epsilon'_1\rangle&=& \left(\begin{array}{c} 1 \nonumber \\ 
0\end{array}\right),\\
\vert\epsilon'_2\rangle&=&\frac{1}{\sqrt{1+\xi^2}} \, \left(\begin{array}{c} 1\\
\xi\, e^{i(\frac{\phi+2k'\pi}{N-2})}  \, \end{array}\right),\ 
k'=0,\,1,\ldots\, N-3 \ \mbox{and}\ \xi=\left(\tan\frac{\theta}{2}\right)^{\frac{1}{N-2}}\\
\vert\epsilon'_3\rangle&=& \left(\begin{array}{c} 0\\
1\, \end{array}\right),\  \nonumber
 \end{eqnarray} 
Here too, $\phi=0$ when $N-2$ is odd and $\phi=\pi$ when $N-2$ is even.
\end{itemize}|

\subsection {Interconvertibility of $3$-qubit GHZ and W$\bar{\rm W}$ states under local operations }
  
If the states (\ref{symghz}), (\ref{symww}) are related to each other by identical local operations of the form $A\otimes A\otimes A,$ i.e., 
\begin{equation}
\label{wtoghz}
\vert {\rm GHZ}\rangle_3={\cal N}\, \left(A\otimes A\otimes A\right)\,\vert {\rm W\bar W} \rangle_3
\end{equation}
  the corresponding Majorana 
spinors (\ref{ghzsp}) and (\ref{wsp}) must be related to each other through 
$\vert\epsilon_r\rangle \,\propto  A\vert \, \epsilon'_s\rangle$, $r,\,s=1,\,2,\,3$; $\cal N$ is the normalization constant.
We consider an arbitrary $2\times 2$ invertible matrix $A=\left(\begin{array}{cc}a & b \\ c & 
d\end{array}\right),\ \ \det A=ad-bc=1$ and explicitly solve the following equations relating Majorana spinors 
(\ref{ghzsp}) and (\ref{wsp}): 
\begin{eqnarray}
\vert\epsilon_1\rangle =N_1\,  A\, \vert\epsilon'_1\rangle \Rightarrow   &&\frac{1}{\sqrt{2}}\,  \left(\begin{array}{c}1\\
\,1 \end{array}\right)=N_1\, \left(\begin{array}{cc}a & b \\ c & d\end{array}\right)
 \left(\begin{array}{c} 1 \\
\,0 \end{array}\right) \nonumber \\ 
\vert \epsilon_2 \rangle =N_2\,  A\, \vert \epsilon'_2\rangle \Rightarrow   
&&\frac{1}{\sqrt{2}}\,  \left(\begin{array}{c}1\\
\,\omega^2 \end{array}\right)=N_2\, \left(\begin{array}{cc}a & b \\ c & d\end{array}\right)
\frac{1}{\sqrt{2}}\,  \left(\begin{array}{c} 1 \\
\,1 \end{array}\right) \nonumber \\   
\vert \epsilon_3\rangle=N_3\,  A\, \vert\epsilon'_3\rangle \Rightarrow  && \frac{1}{\sqrt{2}}\,  \left(\begin{array}{c}1\\
\,\omega \end{array}\right)=N_3\, \left(\begin{array}{cc}a & b \\ c & d\end{array}\right)
  \left(\begin{array}{c} 0 \\
\,1 \end{array}\right)   
\end{eqnarray}
It can be seen that the  SL(2,C) matrix 
\begin{equation}
\label{A}
A=\frac{1}{\omega(\omega-1)}\, \left(\begin{array}{cc} 1 & \omega \\ 1 & \omega^2\, \end{array}\right),
  \end{equation} 
	and the proportionality constants 
	\begin{equation}
\label{N123}
N_1=\sqrt{2},\ \ N_2=-\omega^2,\ \ N_3=\omega\sqrt{2}
  \end{equation} 
lead to the transformation (\ref{wtoghz}). In other words, the $3$-qubit W$\bar{\rm W}$ state (\ref{symww}) gets transformed to the $3$-qubit 
GHZ state (\ref{symghz}) under identical local operations (\ref{A}) on all the three qubits.

\section{Pairwise entanglement features of  $N$-qubit W$\bar{\rm W}$  states}

We adopt concurrence~\cite{hill,wootters} to quantify the pairwise entanglement in W$\rm \bar W$ states. The concurrence $c(\rho)$ of any two-qubit state $\rho$ is defined as~\cite{hill,wootters} 
\begin{equation} 
	\label{hw}
	C(\rho)={\rm max}\,\left(0,\,\sqrt{\lambda_1}-\sqrt{\lambda_2}-\sqrt{\lambda_3}-\sqrt{\lambda_4}\right)
\end{equation}
where $\lambda_i$, $i=1,\,2,\,3,\,4$ are the eigenvalues of the $4\times 4$ matrix  $R=\rho\, (\sigma_{y}\otimes\sigma_{y})\, \rho^{*}\,
(\sigma_{y}\otimes\sigma_{y})$ arranged in the decreasing order $\lambda_1\geq\lambda_2\geq \lambda_3\geq \lambda_4$.

In order to evaluate the pairwise entanglement features of $N$-qubit W$\bar{\rm W}$  states, we need to evaluate the structure of its 
two-qubit subsystems. The state $\vert {\rm W\bar W} \rangle_N$  being a symmetric state, 
its two qubit reduced density matrices are all identical and from its structure in the qubit basis (see (\ref{wwbnq})) one can readily evaluate the two-qubit reduced density matrices 
$\rho_2^{(N)}=\mbox{Tr}_{N-2}\,\vert {\rm W\bar W}\rangle_N\langle {\rm W\bar W}\vert$. We denote $R_N=\rho^{(N)}_2\, (\sigma_{y}\otimes\sigma_{y})\, \rho^{(N)*}_2\,
(\sigma_{y}\otimes\sigma_{y})$ associated with the reduced two-qubit system  $\rho_2^{(N)}$ of the $N$-qubit state $\vert {\rm W\bar W} \rangle_N$. 

\begin{itemize}
	\item For the state $\vert {\rm W\bar W}\rangle_3$ (see (\ref{symww})), 
the two qubit reduced density matrix  obtained by tracing out {\emph{ any}} one qubit is given by 
\begin{eqnarray} 
\rho^{(3)}_2&=&\frac{1}{6}\left[\vert 00\rangle \langle 00\vert+\vert 00\rangle \langle 01\vert+\vert 00\rangle \langle 10\vert+
\vert 01\rangle \langle 00\vert+2\vert 01\rangle \langle 01\vert+2\vert 01\rangle \langle 10\vert+\vert 01\rangle \langle 11\vert \right.\nonumber \\
& &\left. +\vert 10\rangle \langle 00\vert+2\vert 10\rangle \langle 01\vert+2\vert 10\rangle \langle 10\vert+\vert 10\rangle \langle 11\vert 
 +\vert 11\rangle \langle 01\vert+\vert 11\rangle \langle 10\vert+\vert 11\rangle \langle 11\vert\right].
\end{eqnarray}
On expressing $\rho^{(3)}_2$ in $4\times 4$ matrix form (in the basis $\{\vert 00\rangle,\vert 
01\rangle, \vert 10\rangle, \vert 11\rangle \}$, we have 
\begin{eqnarray}
\label{ww2q3}
\rho^{(3)}_2&=&\frac{1}{6}\left(\begin{array}{cccc} 
 1& 1 & 1 & 0 \\ 
1 & 2 & 2 & 1 \\ 
1 & 2 & 2 & 1 \\
0 & 1 & 1 & 1
\end{array}\right). 
\end{eqnarray}

\item For the $4$ qubit W\rm ${\bar W}$ state    
\begin{eqnarray}
\vert {\rm {W\bar W}}\rangle_4&=&\frac{1}{4}\left[ \vert 1_1 0_20_30_4\rangle +\vert 0_11_20_30_4\rangle+\vert 0_10_21_30_4\rangle+\vert 0_10_20_31_4\rangle \right.\nonumber \\
& &\left.+ \vert 0_1 1_21_31_4\rangle +\vert 1_1 0_2 1_3 1_4\rangle+\vert 1_1 1_20_31_4\rangle+\vert 1_1 1_21_30_4\rangle\right],\nonumber 
\end{eqnarray}
the two-qubit reduced density matrix $\rho^{(4)}_2$, obtained by tracing over any two-qubits of $\vert {\rm W\bar W}\rangle_4$, is given by 
\begin{eqnarray} 
\rho^{(4)}_2&=&\frac{1}{4}\left[\vert 00\rangle \langle 00\vert+\vert 00\rangle \langle 11\vert+
\vert 01\rangle \langle 01\vert+\vert 01\rangle \langle 10\vert+\vert 10\rangle \langle 01\vert+\vert 10\rangle \langle 10\vert \right.\nonumber \\
& &\left. +\vert 11\rangle \langle 00\vert+\vert 11\rangle \langle 11\vert\right].
\end{eqnarray}
Thus, the $4\times 4$ matrix representation of $\rho^{(4)}_2$ in the standard two-qubit basis is  
\begin{equation}
\label{ww2q4}
\rho^{(4)}_2=\frac{1}{4}\left(\begin{array}{cccc} 
1 & 0 & 0 & 1  \\ 
0 & 1 & 1 & 0\\ 
0 & 1 & 1 & 0 \\
1 & 0 & 0 & 1
\end{array}\right). 
\end{equation}
\item For any $N>4$, tracing out $(N-2)$ qubits of $\vert {\rm W\bar W}\rangle_N$ (see (\ref{wwbnq}))  
leads to the  two qubit reduced density matrix $\rho^{(N)}_2$: 
\begin{eqnarray}
\label{ww2qN}
\rho^{(N)}_2&=&\frac{N-2}{2N}\,\left[  \vert 00\rangle \langle 00\vert +\vert 1 1\rangle \langle 1 1\vert\right]+ 
\frac{1}{N}\, \left[\vert 01\rangle\langle 0 1\vert + \vert  01\rangle\langle 10\vert+\vert 10\rangle\langle 01   \vert + \vert 10\rangle\langle 10\vert\right] \nonumber \\ 
& & \nonumber \\
&=& \left(\begin{array}{cccc} 
\frac{N-2}{2N} & 0 & 0 & 0 \\ 
0 & \frac{1}{N} & \frac{1}{N} & 0 \\ 
0 & \frac{1}{N} & \frac{1}{N} & 0 \\
0 & 0 & 0 & \frac{N-2}{2N}
\end{array}\right). 
\end{eqnarray}

\item On explicit evaluation of the eigenvalues of $R_3=\rho^{(3)}_2\, (\sigma_{y}\otimes\sigma_{y})\, \rho^{(3)*}_2$, we find that  $\lambda_1=1/4$, $\lambda_2=1/36$, $\lambda_3=\lambda_4=0$ for $N=3$. Thus, the concurrence of the state $\rho^{(3)}_2$ is  $C^{(3)}=1/3$. For $N=4$, the matrix $R_N$ has only two non-zero eigenvalues $\lambda_1=\lambda_2=1/4$, implying that the concurrence of the two-qubit reduced system $\rho^{(4)}_2$ of   $\vert{\rm {W\bar W}}\rangle_4$ state is zero. 

\item For any $N>4$, the eigenvalues of $R_N=\rho^{(N)}_2\, (\sigma_{y}\otimes\sigma_{y})\, \rho^{(N)*}_2\,
(\sigma_{y}\otimes\sigma_{y})$ are
\begin{equation}
\label{cn6}
\lambda_1=\lambda_2=\frac{N-2}{2N}, \ \ \ \lambda_3=\frac{2}{N}, \ \ \ \lambda_4=0.
\end{equation}
\item For $N=5,\ 6$, the values of $\lambda_i$, $i=1,\,2,\,3,\,4$ are found to be
\begin{eqnarray}
\label{56}
N=5: & & \  \lambda_1=\lambda_2= \frac{3}{10}, \ \lambda_3=\frac{2}{5}, \  \lambda_4=0 \nonumber \\
& & \nonumber \\ 
N=6: & & \ \lambda_1=\lambda_2= \frac{1}{3},\  \lambda_3= \frac{2}{3},\  \lambda_4=0. 
\end{eqnarray} 
\item It can be seen from (\ref{56}) that $\sqrt{\lambda_1}-\sqrt{\lambda_2}-\sqrt{\lambda_3}-\sqrt{\lambda_4}$ turns out to be negative and hence concurrence (see (\ref{hw})) of the W$\bar {\rm W}$ state vanishes when $N=5,\,6$. 

\item  From (\ref{cn6}), we have  $\lambda_{1,\,2}>\lambda_3$ when $N>6$. The highest eigenvalue $\lambda_1=\lambda_2$ being doubly repeated, concurrence of the 
state $\vert {\rm W\bar{W}}\rangle_N$ vanishes for $N>6$. In other words,  except for the reduced two-qubit systems $\rho^{(3)}_2$ of the 
the $3$-qubit state $\vert {\rm W\bar{W}}\rangle_3$, the concurrence  for $\rho^{(N)}_2$ drawn from $\vert {\rm W\bar{W}}\rangle_N$  vanishes when $N>3$. Thus,  there is no pairwise entanglement in $\vert {\rm W\bar{W}}\rangle_N$ state except when $N=3$. 
\end{itemize}

It is well-known that there is a trade-off between  entanglement shared by multiple parties. We explore these restrictions on shareability  or  monogamy of entanglement  in $N$-qubit W$\bar{\rm W}$ states. The Coffman–Kundu–Wootters inequality for monogamy of concurrence in a pure three-qubit state is given by~\cite{ckw} 
\begin{equation}
	\label{mt}
	C^2_{12}+C^2_{13}\leq C^2_{1(23)}
\end{equation}
where $C_{12},\ C_{13}$ denote concurrences of two-qubit reduced systems $1,2$ and $1,3$  respectively; $C_{1(23)}$ is the concurrence of the bipartition  $1-23$. The $3$-tangle of a three-qubit state 
$\tau_{3}$, defined by~\cite{ckw} 
\begin{equation}
	\label{3t}
\tau_{3}=C^2_{1(23)}-	C^2_{12}-C^2_{13}
\end{equation}
is a measure of {\emph {residual entanglement}}~\cite{pjg}, which is not accounted by the entanglement between two-qubit subsystems of a three-qubit pure state.  The 3-tangle $\tau_{3}=1$ for three-qubit GHZ states and  $\tau_{3}=0$ for W states.

A generalization of the 3-tangle to any $N$-qubit pure states, called $N$-tangle, has also been proposed~\cite{li1, li2}. The $N$-tangle $\tau_N$ for a pure symmetric $N$-qubit state $\vert \Psi\rangle_{\rm sym}$   is given by  
\begin{equation}
\label{Nts}
\tau_N=4\,\mbox{det}\,\rho^{(N)}_1 -(N-1) C^2; \ \ \rho^{N}_1=\mbox{Tr}_{N-1}\, \vert \Psi\rangle_{\rm sym}\langle 
\Psi\vert.      
\end{equation}
\begin{itemize}
\item One can readily see, by tracing out a qubit from the two-qubit density matrices  (\ref{ww2q3}), (\ref{ww2q4})),  that the single qubit density matrices $\rho^{(3)}_1$ 
$\rho^{(4)}_1$ of $\vert {\rm W\bar W}\rangle_3$, $\vert {\rm W\bar W}\rangle_4$ are obtained as,  
\begin{eqnarray} 
\label{sqww3}
\rho^{(3)}_1&=&=\ba{cc} \frac{1}{2} & \frac{1}{3} \\  \frac{1}{3} & \frac{1}{2} \ea \\ 
\label{sqww4}
\rho^{(4)}_1&=&
\frac{1}{2}\ba{cc}  1 & 0 \\ 0 & 1\ea  
\end{eqnarray}
\item Also, by tracing out a single qubit from the two-qubit states $\rho^{(N)}_2$ (see (\ref{ww2qN})), it can be readily seen that the single qubit density matrix  $\rho^{(N)}_1$  of the state $\vert {\rm W\bar W}\rangle_N$  turns out to be $I_2/2$ for any $N\geq 4$, where $I_2$ is the $2\times 2$ identity matrix. 
\item For the $3$-qubit W$\bar{\rm W}$ state, we have concurrence $C=1/3$ and $\det \rho^{(3)}_1=5/36$ (see (\ref{sqww3})). The 3-tangle $\tau_3=4\,\det \rho^{(3)}_1-2\,C^2=1/3.$ 
\item The concurrence $C$ between any pair of qubits in $\vert {\rm W\bar{W}}\rangle_N$ being zero and $\det \rho^{(N)}_1=\det\, I_2/2=1/4$, $N\geq 4$, we obtain the $N$-tangle (see (\ref{Nts})) 
$\tau_N=1$ for  $N\geq 4$. 
\end{itemize}

From the above results, we conclude that two-qubit entanglement in $N$-qubit W$\bar{\rm W}$ states with $N\geq 4$ vanishes - but their residual entanglement quantified by $N$-tangle $\tau_N$ is maximum. The three-qubit  W$\bar{\rm W}$ state possesses both two-qubit entanglement (quantified by concurrence $C=1/3$) and residual 3-way entanglement (characterized by $\tau_{3}^{\rm W\bar W}=1/3$).  


It is important to notice that $N$-qubit GHZ state also has vanishing concurrence and maximum concurrence tangle~\cite{ckw} $\tau=1$ for any $N\geq 3$. The identical pairwise entanglement and monogamy features of GHZ and W$\bar{\rm W}$ states for any $N\geq 4$ is worth noticing. We examine this aspect with the help of  canonical steering ellipsoids  in Sec.V. 

 \section{Canonical steering ellipsoids as geometric representation of two-qubit states} 
It has been shown~\cite{supra,can,pra23} that the SLOCC transformation on a two-qubit state is equivalent to the Lorentz congruent transformation (up to a scalar factor) on the real representative of the state. This helps in obtaining the Lorentz canonical form of the real representative of the state and thereby its geometric representation in terms of canonical steering ellipsoiods~\cite{supra, can,pra23}. While the canonical steering ellipsoids corresponding to two SLOCC inequivalent families of $3$-qubit states are obtained in Ref. \cite{can}, the canonical steering ellipsoids associated with SLOCC inequivalent families of $N$-qubit Dicke class of states  are analysed in Ref. \cite{pra23}.
In the following, after a brief description of the concepts used in~\cite{supra,can, pra23}, we proceed to work on the geometrical representation of $N$-qubit GHZ and W$\rm{\bar W}$ states.  


\subsection{Real representation of two-qubit states and their Lorentz canonical forms} 

Consider a two-qubit density matrix $\rho_2$ expressed in the  Hilbert-Schmidt basis $\{\sigma_\mu\otimes \sigma_\nu\}$:  
\begin{eqnarray}
	\label{rho2q}
	\rho_2&=&\frac{1}{4}\, \sum_{\mu,\,\nu=0}^{3}\,   
	\Lambda_{\mu \, \nu}\, \left( \sigma_\mu\otimes\sigma_\nu \right), 
\end{eqnarray}
The coefficients of expansion $\Lambda_{\mu \, \nu}$ 
\begin{eqnarray}
	\label{lambda}
	\Lambda_{\mu \, \nu}&=& {\rm Tr}\,\left[\rho_2\,
	(\sigma_\mu\otimes\sigma_\nu)\,\right],  
\end{eqnarray}  
form a real $4\times 4$ matrix $\Lambda$ representing the density matrix $\rho_2$.
Here, $\sigma_i$, $i=1,\,2,\,3$ are the Pauli spin matrices and $\sigma_0$ is the $2\times 2$ identity matrix. 

\begin{itemize}	
\item When the qubits are subjected to local operations on their respective parts,  the two-qubit density matrix $\rho_2$ transforms to $\widetilde{\rho}_2$ as  
	\begin{eqnarray}
	\label{rhokab}
		\rho_2\longrightarrow\widetilde{\rho}_2&=&\frac{(A\otimes B)\, \rho_2\, (A^\dag\otimes B^\dag)}
		{{\rm Tr}\left[\rho_2\, (A^\dag\, A\otimes B^\dag\, B)\right]}.
	\end{eqnarray} 
	Here, $A, B\in {\rm SL(2,C)}$ denote $2\times 2$ complex matrices with unit determinant and represent local operations on the qubits $A$, $B$.
	One can choose suitable local operations $A$ and $B$ such that the two-qubit density matrix $\rho_2$ attains its canonical form $\widetilde{\rho}_2$.
	
\item When the two-qubit state $\rho_2$ undergoes the transfomation (\ref{rhokab}), its real representative $\Lambda$ transforms as~\cite{supra,can}  
	\begin{eqnarray}
		\label{sl2c}
		\Lambda\longrightarrow \widetilde{\Lambda}&=&\frac{L_A\,\Lambda\, L^T_B}{\left(L_A\,\Lambda\, L^T_B\right)_{00}}.
	\end{eqnarray} 
Here $L_A,\, L_B\in SO(3,1)$ are $4\times 4$  proper orthochronous Lorentz transformation matrices~\cite{kns} corresponding respectively to  $A$, $B\in SL(2,C)$  and the superscript `$T$' denotes transpose operation.
\item	The Lorentz canonical form~\cite{supra} $\widetilde{\Lambda}$ can be obtained by constructing the $4\times 4$ real symmetric matrix $\Omega=\Lambda\, G\, \Lambda^T$, where $G={\rm diag}\,(1,-1,-1,-1)$ denotes the Lorentz metric~\cite{kns}. 
\item Using the defining property~\cite{kns}   $L^T\,G\,L=G$  of Lorentz transformation $L$,  it can be seen that the matrix $G\Omega$
	undergoes a {\emph similarity transformation} $G\widetilde \Omega=L^{-1} (G\Omega)L$.   
\item It has been shown~\cite{supra,can,pra23} that the canonical form of 
$\Lambda$ can either be a diagonal matrix or a non-diagonal matrix with only one off-diagonal element~\cite{verstraete2001, supra,can,pra23}, depending on the eigenvalues and eigenvectors of $G\Omega$. \item When the eigenvector $X_0$ associated with the highest eigenvalue $\lambda_0$ of $G\,\Omega$ obeys the   Lorentz invariant condition
		$X_0^T\, G\, X_0>0$, $\Lambda$ assumes the diagonal canonical form $\widetilde\Lambda_{\rm I}$ given by
			\begin{eqnarray} 
			\label{lambda1c}
			\widetilde\Lambda_{\rm I}&=&{\rm diag}\,  \left(1,\,\sqrt{\frac{\lambda_1}{\lambda_0}},\sqrt{\frac{\lambda_2}{\lambda_0}},\, \pm\, \sqrt{\frac{\lambda_3}{\lambda_0}}\right), 
			\end{eqnarray} 
		where $\lambda_0\geq\lambda_1\geq\lambda_2\geq \lambda_3> 0$ are the  {\em non-negative} eigenvalues of $G\,\Omega$. 	 
\item Suppose that the non-negative eigenvalues $\lambda_0$ and $\lambda_1$ ($\lambda_0\geq \lambda_1$) of $G\Omega$ are doubly degenerate and an eigenvector $X_0$ of $G\Omega$, belonging to the highest eigenvalue $\lambda_0$, satisfies the Lorentz invariant condition 	
$X_0^T\, G\, X_0 =0$. In such cases,	the Lorentz canonical form of $\Lambda$ turns out to be a non-diagonal matrix (with only one non-diagonal element):  
\begin{eqnarray} 
			\label{lambda2c}
			\widetilde{\Lambda}_{\rm II}&=& \left(\begin{array}{cccc}
				1 & 0  & 0 & 0 \\ 
				0 & a_1 & 0 & 0 \\ 
				0 & 0 &   -a_1 & 0 \\
				1-a_0 & 0 & 0 &  a_0 \ \ 
			\end{array}\right) \ \ 
			\end{eqnarray}
The parameters $a_0$, $a_1$ in (\ref{lambda2c}) are given by~\cite{supra,can,pra23} 
\begin{eqnarray}
\label{phi0}
&& a_0=\frac{\lambda_0}{\phi_0},\ \ a_1=\sqrt{\frac{\lambda_1}{\phi_0}}
\end{eqnarray} 
where $\phi_0$ is the $00^{\rm th}$ element of the canonical form $\widetilde {\Omega}_{\rm II}={\widetilde{\Lambda}_{II}}\,G\,{\widetilde{\Lambda}}^T_{\rm II}$: 
\begin{eqnarray}
	\label{yyy}
	\widetilde{\Omega}_{II}&=&\,\left(\begin{array}{cccc}
				\phi_0 & 0  & 0 & \phi_0-\lambda_0 \\ 
				0 & -\lambda_1 & 0 & 0 \\ 
				0 & 0 &   -\lambda_1 & 0 \\
				\phi_0-\lambda_0 & 0 & 0 &  \phi_0-2\lambda_0 \ \ 
			\end{array}\right).
	\end{eqnarray} 
\end{itemize}
\subsection{Steering ellipsoids corresponding to Lorentz canonical form of two-qubit states}
In the two-qubit state $\rho_2$, local projective valued measurements (PVM)
\begin{equation}
	Q=I_2 + \sigma_1 q_1 + \sigma_2 q_2 + \sigma_3 q_3,\ \ q_1^2+q_2^2+q_3^2=1 
\end{equation}
on Bob's qubit  leads to  collapsed states of Alice's qubit characterized by  Bloch vectors 
${\mathbf p}_A=(p_1,\,p_2,\,p_3)^T$ through the transformation~\cite{supra} 
\begin{equation} 
\label{funda}
\left(1, p_1,\,p_2,\,p_3 \right)^T=\Lambda\,\left(1, q_1,\,q_2,\,q_3 \right)^T, \ \ q_1^2+q_2^2+q_3^2=1. 
 \end{equation}
Here  ${\mathbf{q}}_B=\left(q_1,\,q_2,\,q_3 \right)^T$, $q_1^2+q_2^2+q_3^2=1$ represents points on the Bloch sphere representing all possible PVMs at Bob's end.  The steered Bloch vectors ${\mathbf p}_A$ of Alice's qubit constitute an ellipsoidal surface ${\mathcal E}_{A\vert\,B}$, enclosed within the Bloch sphere. 

\begin{itemize}
\item For the diagonal  canonical form ${\widetilde{\Lambda}_{I}}$  (see (\ref{lambda1c})) of the two-qubit state, it follows from (\ref{funda}) that
\begin{equation}
		p_1=\sqrt{\frac{\lambda_1}{\lambda_0}}\,q_1, \ \ p_2=\sqrt{\frac{\lambda_2}{\lambda_0}}\,q_2,\ \ p_3=\pm \sqrt{\frac{\lambda_3}{\lambda_0}} q_3, \ \ 
					\end{equation} 
are steered Bloch points ${\mathbf {p}}_A$ of Alice's qubit.  They  obey the equation 
		\begin{equation}
		\label{ellI}
		\frac{\lambda_0\, p_1^2}{\lambda_1}+ \frac{\lambda_0\, p_2^2}{\lambda_2}+ \frac{\lambda_0\, p_3^2}{\lambda_3}=1
		\end{equation}
		of an  ellipsoid with semiaxes  $a_1=\sqrt{\lambda_1/\lambda_0}, \,a_2=\sqrt{\lambda_2/\lambda_0},\, a_3=\sqrt{\lambda_3/\lambda_0})$ and center $(0,0,0)$ inside the Bloch sphere $q_1^2+q_2^2+q_3^2=1$. We refer to this as the {\em canonical steering ellipsoid} representing the set of all two-qubit density matrices which are SLOCC equivalent to the canonical form $\widetilde{\rho}_{\rm I}$ (see (\ref{rhokab})) corresponding to ${\widetilde{\Lambda}_{I}}$. 	

\item For the non-diagonal canonical form $\widetilde{\Lambda}_{\rm II}$ (see (\ref{lambda2c})), we get the coordinates of steered Alice's Bloch vector ${\mathbf{p}}_A$, on using (\ref{funda});
\begin{equation}
		\label{e2A}
	p_1=a_1 q_1,\ \  p_2=-a_1q_2,\ \ p_3= \left(1-a_0\right)+a_0q_3, \   \ \ q_1^2+q_2^2+q_3^2=1
			\end{equation} 
and they satisfy the equation 
\begin{eqnarray}
		\label{sph}
	&& \frac{p_1^2}{a_1^2}+ \frac{p_2^2}{a_1^2}+ \frac{\left(p_3-(1-a_0)\right)^2}{a_0^2}=1.	
		\end{eqnarray}  

Note that (\ref{sph}) represents the canonical steering spheroid (traced by Alice's Bloch vector ${\mathbf{p}}_A$) inside the Bloch sphere with its center at $(0,\,0,\, 1-a_0)$ and lengths of the semiaxes given by $a_0=\lambda_0/\phi_0$, $a_1=\sqrt{\lambda_1/\phi_0}$. In other words, a shifted spheroid inscribed within the Bloch sphere, represents  two-qubit states that possess a non-diagonal Lorentz canonical form $\widetilde{\Lambda}_{\rm II}$ (see (\ref{lambda2c})). 
\end{itemize}
In the following, we obtain the Lorentz canonical forms of the real representation $\Lambda^{(N)}$ associated with the two-qubit subsystems $\rho^{(N)}_{2}$ of $N$-qubit W$\rm{\bar W}$ states for all $N$ and arrive at their canonical steering ellipsoids.

\section{Canonical steering ellipsoids corresponding to $\vert{\rm  W\bar{W}}\rangle_N$} 
In Sec. II, we have seen that the $N$-qubit ${\rm  W\bar{W}}$ ($N>3$) states do not have pairwise entanglement (zero concurrence) and exhibit maximum restriction in the shareability of entanglement among its subsystems (concurrence tangle $\tau=1$). In contrast to this, the 
$3$-qubit ${\rm W\bar W}$ state have non-zero concurrence ($C=1/3$) and the concurrence tangle is not maximum ($\tau=1/3$). 
 In this section, we explore how the geometric picture of the two-qubit subsystems of the states   $\vert {\rm  W\bar{W}}\rangle_N$ reflect this aspect.

 \subsection{Canonical steering ellipsoids corresponding to $\vert{\rm  W\bar{W}}\rangle_3$} 
The real matrix representation $\Lambda_3$ of the two-qubit subsystem $\rho^{(3)}$ (see (\ref{ww2q3})) of  $\vert {\rm W\bar{W}}\rangle_3$ and the real symmetric matrix $\Omega^{(3)}=\Lambda_3\,G\,\left(\Lambda_3\right)^T$ are respectively given by 
\begin{eqnarray} 
\label{l3}
\Lambda_3&=&\ba{cccc} 1 & \frac{2}{3} & 0 & 0 \\ \frac{2}{3} & \frac{2}{3} & 0 & 0 \\ 0 & 0 & \frac{2}{3} & 0 \\ 0 & 0 & 0 & \frac{1}{3} \ea, \ \ \mbox{and} \ \ 
\Omega^{(3)}=\frac{1}{9}\ba{rccc} 5 & 2 & 0 & 0 \\ -2 & 0 & 0 & 0 \\ 0 & 0 & 4 & 0  \\  0 & 0 & 0 & 1\ea.
\end{eqnarray} 
The normalized eigenvectors $X_k$ of $G\Omega^{(3)}$ belonging to its eigenvalues $\lambda_1=\lambda_2=4/9$, $\lambda_3=\lambda_4=1/9$ can be determined by solving the eigenvalue equation $\left(G\Omega^{(3)}\right)X_k=\lambda_k\,X_k$, $k=0,\,1,\,2,\,3.$ On explicit determination, we get 
\begin{eqnarray} 
X_0&=&\frac{2}{\sqrt{3}}\left(2,\,-1,\,0,\,0\right)\in \lambda_0=4/9 \\
X_1&=&\left(0,\,1,\,0,\,0\right)\in \lambda_1=4/9 \nonumber \\ 
X_2&=&\frac{1}{\sqrt{3}}\left(1,\,-2,\,0,\,0\right)\in \lambda_2=1/9 \nonumber \\ 
X_3&=&\left(0,\,0,\,0,\,1\right)\in \lambda_3=1/9 \nonumber 
\end{eqnarray}
It can be readily verified that $X_r^T\,G\,X_s=g_{rs}$ where $g_{00}=1$, $g_{11}=g_{22}=g_{33}=-1$, 
$g_{rs}=0$ when $r\neq s$, $r,\,s=0,\,1,\,2,\,3$ i.e.,  $X_0^T\,G\,X_0=1$ and $X_k^T\,G\,X_k=-1$ when $k=1,\,2,\,3$. Thus, the set $\left\{X_0,\,X_1,\,X_2,\,X_3\right\}$ forms an orthonormal tetrad of Minkowski four-vectors~\cite{kns}. The $4\times 4$ real matrix 
$L=\left(X_0,\, X_1,\, X_2,\, X_3\right)$, constructed using the eigenvectors of $G\Omega^{(3)}$ is a Lorentz matrix~\cite{kns} satisfying the relation $L^T\,G\,L=G$ and $\mbox{det}\,L=1$. Explicitly, we have 
\be
L=\frac{1}{\sqrt{3}}\ba{rrcc} 2 & -1 & 0 & 0 \\ 0 & 1 & 0 & 0 \\ -1 & 2 & 0 & 0 \\ 0 & 0 & 0 & 1 \ea.
\ee 
From the relations $\left(G\Omega^{(3)}\right)X_k=\lambda_k\,X_k$, $X_r^T\,G\,X_s=g_{rs}$, one can readily see that $\widetilde \Lambda_3=L \Lambda_3 L^T$ is a diagonal matrix with elements $a_k=\sqrt{\frac{\lambda_k}{\lambda_0}}$. With $\lambda_1=\lambda_2=4/9$, $\lambda_3=\lambda_4=1/9$, we have $a_1=1$, $a_2=a_3=1/2$. 

From the discussion in Sec. IVB leading to (\ref{ellI}), 
it can be seen that the canonical steering ellipsoid corresponding to $\vert{\rm  W\bar{W}}\rangle_3$ is an oblate spheroid  with radius $1/2$ centered at the origin and touching the Bloch sphere (see Fig.~1)
\begin{figure}[h]
	\begin{center}
		\includegraphics*[width=3in,keepaspectratio]{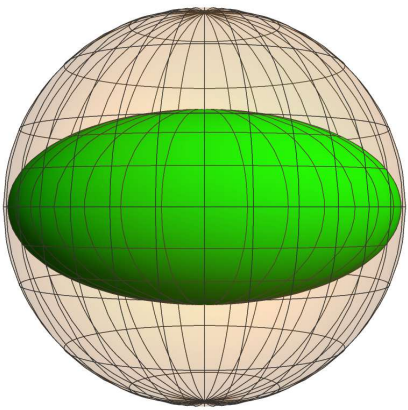}
		\caption{Oblate spheroid representing the state $\vert{\rm  W\bar{W}}\rangle_3$: Length of the semi-axes: $a_1=1$, $a_2=a_3=1/2$}
	\end{center}
\end{figure}  

\subsection{Geometrical representation of $4$-qubit W${\rm\bar W}$ states} 

The two-qubit subsystem $\rho^{(4)}_2$  of the $4$-qubit $\rm{W\bar W}$ state given in (\ref{ww2q4}) leads to the diagonal structure of its real representative $\Lambda_4$;
\be
\label{l4}
\Lambda_4=\mbox{diag} \left(1,\,1,\,0,\,0\right)
\ee 
Thus $G\Omega^{(4)}=G\left(\Lambda_4G(\Lambda_4)^{T}\right)=\Lambda_4$ and it is already in its canonical form. The eigenvector belonging to the largest eigenvalue $\lambda_0=\lambda_1=1$ is $X_0=\left(1,\,0,\,0,\,0\right)$ of $G\Omega^{(4)}$ and satisfies the relation
$\left(X_0\right)^T\,G\,X_0=1$. With the other two eigenvalues of $G\Omega^{(4)}$ being $\lambda_2=\lambda_3=0$, 
the parameters $a_1=\sqrt{\lambda_1/\lambda_0}$, $a_2=a_3=\sqrt{\lambda_{2,\,3}/\lambda_0}$ are 
$a_1=1$, $a_2=a_3=0$ i.e.,  the tip of Bloch vectors of one  qubit, steered by another qubit of the $4$-qubit $\rm{W\bar W}$ state trace a straight line joining the north and south poles of the Bloch sphere. 

It is of interest to note here that the $N$-qubit GHZ states have the same geometrical representation as that of 
$4$ qubit  W${\rm\bar W}$ states. More specifically, the two-qubit subsystem density matrix of $\vert \rm{GHZ}\rangle_N$  is given by
$\rho_2^{\rm GHZ}=\mbox{diag}\left(1/2,\,0,\,0,\,1/2 \right)$ and  its real representative is found to be  
\be
\Lambda_N^{\rm GHZ}=\mbox{diag} \left(1,\,0,\,0,\,1\right). 
\ee 
It readily follows that 
$G\,\Omega^{(N)}_{\rm GHZ}=G\left(\Lambda_N^{\rm GHZ}G(\Lambda_N^{\rm GHZ})^{T}\right)=\Lambda_N^{\rm GHZ}$.
The eigenstructures of $G\Omega$ corresponding to $N$-qubit GHZ states and $4$-qubit W${\rm\bar W}$ state are identical. The two-qubit subsystem density matrix of $N$-qubit GHZ state is represented by a straight line joining north and south poles of the Bloch sphere.

\subsection{Canonical steering ellipsoids corresponding to $N$-qubit W${\rm\bar W}$ states; $N>4$} 

In order to obtain the geometrical representation of $N$-qubit W${\rm\bar W}$ states for $N>4$, we first evaluate the real representative $\Lambda_N$ of two-qubit subsystem $\rho^{(N)}_2$ obtained in (\ref{ww2qN}): We find that  
$$\Lambda_N={\rm diag}\,(1,\ 2/N,\ 2/N,\ (N-4)/N).$$ 
 Eigenvalues of $G\Omega^{(N)}=G\,\Lambda_N\,G\,(\Lambda_N)^{T}$  are given by  
$$\lambda_0=1,\ \lambda_1=\lambda_2=4/N^2,\ \lambda_3=(N-4)^2/N^2$$ and the eigenvector $$X_0=\left(1,\,0,\,0,\,0\right)$$ corresponding to the highest eigenvalue $\lambda_0=1$ satisfies the relation
$X_0^T\,G\,X_0=1$. Thus, the semi-axes of the steering ellipsoid  (see Sec. IVB) corresponding to two-qubit subsystem density matrix $\rho^{(N)}_2$ of 
$\vert\rm {W\bar W}\rangle_N$ are given by
\be
\label{san}
a_1=\sqrt{\frac{\lambda_1}{\lambda_0}}=\frac{2}{N}=a_2;\ \  a_3=\sqrt{\frac{\lambda_3}{\lambda_0}}=\frac{N-4}{N} 
\ee
Thus, the $N$-qubit W${\rm\bar W}$ state is represented geometrically by a spheroid ($a_1=a_2=2/N$)  centered at the origin of the Bloch sphere. For $N=5$, we obtain an oblate spheroid (see Fig. 2).     
\begin{figure}[ht]
	\begin{center}
		\includegraphics*[width=3in,keepaspectratio]{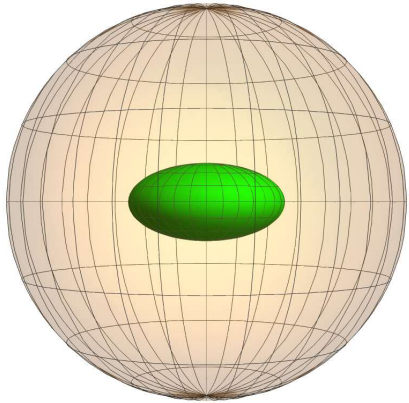}
		\caption{Canonical steering ellipsoid representing $\vert \rm{W\bar{W}}\rangle_5$ with semiaxes  $a_1=a_2=2/5$, $a_3=1/5$}
	\end{center}
\end{figure} 
When $N=6$, it is readily seen from (\ref{san}) that we have a sphere of radius $1/3$ (see Fig. 3).
\begin{figure}[h]
	\begin{center}
		\includegraphics*[width=3in,keepaspectratio]{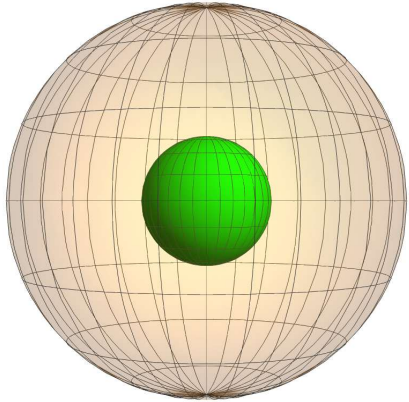}
		\caption{Sphere of radius $\frac{1}{3}$, the canonical steering ellipsoid of $\vert \rm{W\bar{W}}\rangle_6$}
	\end{center}
\end{figure} 
For $N\geq 7$, the canonical steering ellipsoids of $\vert \rm{W\bar{W}}\rangle_N$ are prolate spheroids with their radius $2/N$ decreasing with $N$ (see Fig. 4 and Fig. 5)

\begin{figure}[h]
	\begin{center}
		\includegraphics*[width=3in,keepaspectratio]{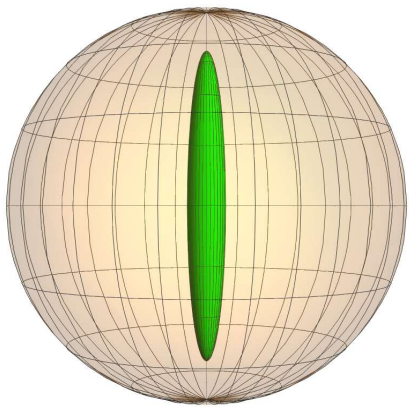}
		\caption{Prolate spheroid with semiaxes $a_1=a_2=\frac{1}{10}$, $a_3=\frac{4}{5}$ representing 
		$\vert {\rm W\bar W}\rangle_{20}$}
	\end{center}
\end{figure} 

\begin{figure}[h]
	\begin{center}
		\includegraphics*[width=3in,keepaspectratio]{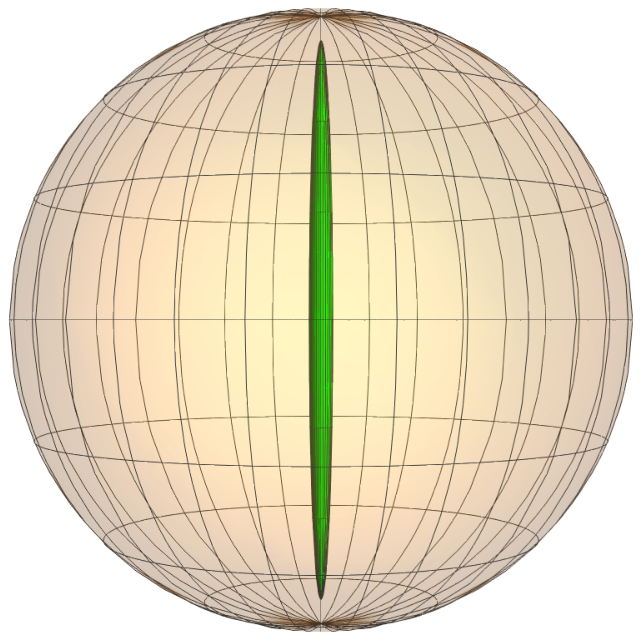}
		\caption{Prolate spheroid with semiaxes $a_1=a_2=1/25$, $a_3=23/25$ representing $\vert {\rm W\bar W}\rangle_{50}$}
	\end{center}
\end{figure} 

One can obtain similar canonical steering ellipsoids for generalized counterparts of GHZ and ${\rm W\bar W}$ states by constructing the real representatives of their two-qubit subsystems. We notice that the nature of the canonical ellipsoids remain the same as that for $\vert \rm{W\bar W}\rangle_N$ and $\vert \rm{GHZ}\rangle_N$ but the volume of the ellipsoid increases when $\theta<\pi/2$, $\theta>\pi/2$. This indicates the increase in shareability of entanglement among its subsystems.  Quite in accordance with increase in pairwise entanglement (concurrence) and decrease in monogamous nature (decrease in concurrence tangle) of these states (see Sec. III), their shareability of entanglement decreases when $\theta>\pi/2$, $\theta<\pi/2$ as reflected through the volume of the corresponding canonical steering ellipsoids.

\section{Volume monogamy features of $\vert {\rm W\bar{W}}\rangle_N$ and $\vert \rm{GHZ}\rangle_N$} 
The restriction on shareability of quantum correlations in a multipartite state get captured in various monogamy relations, which find interesting applications in ensuring security in quantum key distribution~\cite{Tehral,Paw}. A geometrically intuitive monogamy relation in terms of the volumes	of the steering ellipsoids representing the two-qubit subsystems of multiqubit pure states has been proposed and extensively studied  in Refs. \cite{MilneNJP2014,shi2011,MilnePRA2014,shi}. The volume monogamy relation is stronger than the well-known  Coffman-Kundu-Wootters monogamy relation (\ref{mt}) and is applicable to $N$-qubit pure states.

The normalized volume $v_N$  of the quantum steering ellipsoid corresponding to pure symmetric $N$-qubit state $\vert \Psi\rangle_{\rm sym}$is given by~\cite{pra23} 
\begin{equation} 
\label{volN} 
v_N=\frac{\vert\det{\Lambda_N}\vert}{(1-r^2)^2}
\end{equation} 
Here $\Lambda_N$ is the real representative of 
the two-qubit density matrix $\rho^{(N)}_2$ of  $\vert \Psi\rangle_{\rm sym}$ and $\vec{r}=(r_1,\,r_2,\,r_3)$ ($\vert\vec{r}\vert\leq 1$) is the Bloch-vector of the  single qubit  subsystem.  
It has been shown  in Ref.~\cite{pra23}  that the volume monogamy relation in the state $\vert \Psi\rangle_{\rm sym}$ is given by  
\begin{equation}
\label{volm}
\left(v_{N}\right)^{\frac{2}{3}} \leq \frac{1}{2}. 
\end{equation} 
We proceed to study the volume monogamy relation governing $N$-qubit ${\rm W\bar W}$ state.  The LHS of the monogamy inequality 
(\ref{volm}) is a measure of the degree of restriction on shareability of quantum correlations in any arbitrary $N$-qubit pure symmetric state. The lowest possible value $0$ indicates maximum restriction on shareability of entanglement. Equality sign in ({volm})  indicates least restriction and largest allowed shareability of entanglement.    
\begin{figure}[h]
	\begin{center}
		\includegraphics*[width=3in,keepaspectratio]{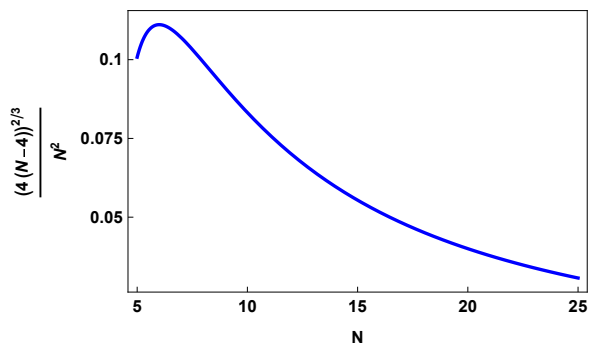}
		\caption{ LHS. of the volume monogamy relation  (\ref{lhs1}) obeyed by $\vert\rm{ W\bar{W}}\rangle_N$ state as a function of the number of qubits $N$.}.
	\end{center}
\end{figure} 

\begin{figure}[h]
	\begin{center}
		\includegraphics*[width=3in,keepaspectratio]{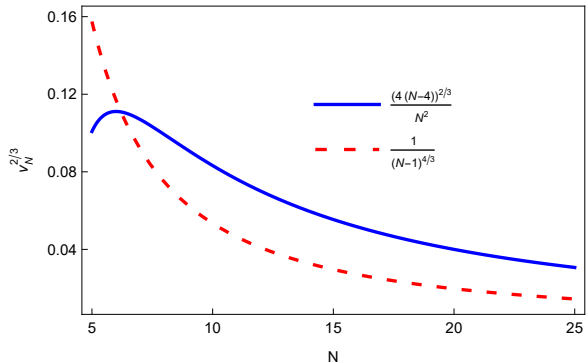}
		\caption{Comparison of the LHS of the volume monogamy relation governing $\vert \rm {W\bar{W}}\rangle_N$ and the W-class of $N$-qubit states.}
	\end{center}
\end{figure} 

For $N=3$,  the real representative $\Lambda_3$ of the two-qubit subsystem $\rho^{\rm W\bar W}_2$ of $\vert {\rm W\bar{W}}\rangle_3$ is given in (\ref{l3}).  It is seen that the Bloch vector components  
$r_1=2/3$, $r_2=r_3=0$ (see (\ref{l3})) leading to  $r=\sqrt{r_1^2+r_2^2+r_3^2}=2/3$. We find that $\vert \det \Lambda^{(3)}\vert=4/81$. Thus we obtain (see (\ref{volN}))  
$v_{3}=4/25$. Thus, volume monogamy relation is readily satisfied: 
$$\left(v_{3}\right)^{2/3}=0.16<1/2.$$

For $\vert {\rm W\bar{W}}\rangle_4$, we obtain $\det \Lambda_4=0$ as $\Lambda_4=\mbox{diag}(1,\,1,\,0,\,0)$ 
(see (\ref{l4})). The volume monogamy relation is thus maximally satisfied. It may be noted that the real matrix  corresponding to two-qubit subsystem of $N$-qubit GHZ state is given by $\Lambda_{\rm{GHZ}}=\mbox{diag}(1,\,0,\,0,\,1)$ and hence, $\det \Lambda_{\rm{GHZ}}=0$ irrespective of the value $N$. Thus, the LHS of the volume monogamy inequality (\ref{volm}) takes its minimum value 0 for GHZ state, highlighting the strongest restrictions on sharability of entanglement.    

Let us turn our attention to volume monogamy property of  $\vert {\rm W\bar{W}}\rangle_N$, $N>4$.  The real representative $\Lambda_N$ of $\vert {\rm W\bar{W}}\rangle_N$, $N>4$ is a diagonal matrix $\Lambda^{(N)}=\mbox{diag}\left(1,\,2/N,\,2/N,\,(N-4)/n\right)$  
with the magnitude of single qubit Block vector $r=\sqrt{r_1^2+r_2^2+r_3^2}=0$. Substituting $\det \Lambda_N=4(N-4)/N^3$ (see (\ref{volN}))   we obtain 
 the volume monogamy relation (\ref{volm}) for $\vert {\rm W\bar{W}}\rangle_N$, $N>4$: 
\begin{equation}
\label{lhs1} 
v_{N}^{2/3}=\frac{\left(4(N-4)\right)^{2/3}}{N^2}\leq \frac{1}{2}.
\end{equation}
We have plotted the LHS of volume monogamy relation (\ref{lhs1}) in Fig.~6. It is evident from Fig.~6 that the sharability of entanglement gets restricted as number of qubits $N$ increase and is quantified by reducing volume of the canonical steering ellipsoids  On explicit evaluation of the volumes of steering ellipsoids (as a function of the parameter $\theta$) of generalized ${\rm W\bar W}$ and GHZ states, we find analogous features. 

It is worth recalling here that the volume monogamy inequality for   W   state belonging to the family ${\cal{D}}_{N-1,1}$ -- which are constituted by two distinct spinors -- is shown~\cite{pra23} to be $(N-1)^{-4/3}<1/2$. A comparison of the LHS of monogamy relation of $\vert \rm {W\bar{W}}\rangle_N$ and the $W$-class of states is depicted  in Fig.~7. It is seen that W states exhibit stricter volume monogamy for pairwise entanglement  than  W$\bar{\rm W}$ states beyond $N\geq 7$ when the cross-over happens (see Fig.~7).

\section{Summary} 
In this work, we have studied the $N$-qubit GHZ and equal superposition of $N$-qubit W, $\rm{\bar W}$ states, which are constituted by $N$-distinct Majorana spinors. We have evaluated the explicit structure of Majorana spinors of both the families. Using the structure of Majorana spinors for $3$-qubit GHZ and $\rm {W\bar W}$ states, we have shown that they are interconvertible under identical local operations on their qubits. Geometric visualization of $N$-qubit $\rm {W\bar W}$ states in terms of   canonical steering ellipsoids inscribed within the Bloch sphere is highlighted. Furthermore, we have investigated  the volume monogamy inequality governing shareability of entanglement in $N$-qubit ${\rm W\bar W}$ states.

\section*{Acknowledgements}
ARU and Sudha  are supported by the Department of Science and Technology (DST), India through Project No. DST/ICPS/QUST/2018/107. ASH is supported
by the Foundation for Polish Science (IRAP Project, ICTQT, contract no. MAB/2018/5, co-financed by EU within Smart Growth Operational Programme). HSK is supported by the Institute of Information \& Communications Technology Planning \& 14 Evaluation (IITP) Grant funded by the Korean government (MSIT) (No.2022-0-00463, Development of a quantum repeater in optical fiber networks for quantum internet).


\begin{thebibliography}{1}
\bibitem{fn} The $N$th roots of any complex number $z=e^{i\phi}$
are given by $z_k=e^{i\phi_k}$ where $\phi_k=\left(\frac{\phi+2k\pi}{N}\right)$, $k=0,1,\,2\ldots N-1$
\end{thebibliography}
\end{document}